\documentclass[%
superscriptaddress,
preprint,
 amsmath,amssymb,
 aps,
]{revtex4-1}

\usepackage{graphicx}
\usepackage{dcolumn}
\usepackage{bm}

\usepackage{color,soul}
\soulregister\cite7
\soulregister\ref7
\usepackage{footnote}

\begin{document}

\title{Hybrid functional pseudopotentials}

\author{Jing Yang}
 \affiliation{Department of Chemistry, University of Pennsylvania, Philadelphia,
 Pennsylvania  19104-6323, USA}
\author{Liang Z. Tan}
 \affiliation{Department of Chemistry, University of Pennsylvania, Philadelphia,
 Pennsylvania  19104-6323, USA}
\author{Andrew M. Rappe}
 \affiliation{Department of Chemistry, University of Pennsylvania, Philadelphia,
 Pennsylvania  19104-6323, USA}
 \email{rappe@sas.upenn.edu}

\begin{abstract}
The consistency between the exchange-correlation functional used in pseudopotential construction and in the actual density functional theory calculation is essential for the accurate prediction of fundamental properties of materials. However, routine hybrid density functional calculations at present still rely on GGA pseudopotentials due to the lack of hybrid functional pseudopotentials. Here, we present a scheme for generating hybrid functional pseudopotentials, and we analyze the importance of  pseudopotential density functional consistency for hybrid functionals. We benchmark our PBE0 pseudopotentials for structural parameters and fundamental electronic gaps of the G2 molecular dataset and some simple solids. Our results show that using our new PBE0 pseudopotentials in PBE0 calculations improves agreement with respect to all-electron calculations.
\end{abstract}

\pacs{Valid PACS appear here}
\maketitle

\section{Introduction}
Density functional theory (DFT) methods have proven to be successful for understanding and predicting the physical and chemical properties of materials. 
With  approximations such as the local density approximation (LDA)~\cite{Kohn65pA1133} and generalized-gradient approximation (GGA)~\cite{Perdew96p3865}, DFT can reproduce many fundamental properties of solids, such as lattice constants and atomization energies~\cite{Buhl08p1449}. However, LDA and GGA usually underestimate the fundamental band gaps of semiconductors and insulators~\cite{Perdew85p497}. The use of hybrid functionals in DFT, which combine part of the exact Hartree-Fock (HF) exchange with local or semilocal approximations (PBE0, HSE, B3LYP)~\cite{Adamo99p6158,Muscat01p397,Heyd05p174101}, has become a popular option for addressing this problem. 

The pseudopotential approximation is often used to reduce the complexity of DFT calculations. By replacing the nucleus and core electrons with a finite shallow potential, the solution of the Kohn-Sham equation is simplified because of the reduced number of electrons in the system. Accuracy is preserved because the core electrons are not involved in chemical bonding~\cite{Rappe90p1227,Ramer99p12471}. 

Even though hybrid density functional calculations using pseudopotentials are currently very popular, these calculations solve the Kohn-Sham equation using pseudopotentials constructed at a lower rung of Jacob's ladder~\cite{Perdew01p1}, such as GGA.
This is due to a lack of hybrid functional pseudopotentials available to the community. The mismatch of the level of density functional approximation between pseudopotential construction and target calculation is theoretically unjustified, and could lead to reduced accuracy~\cite{Fuchs98p2134}. 
In this work, we have developed hybrid density functional pseudopotentials to restore pseudopotential consistency in hybrid functional DFT calculations. 


Prior to this work, Hartree-Fock pseudopotentials developed over the last decade~\cite{Trail05p014112,Al-Saidi08p075112} have proven to be useful in calculations with correlated electrons. The inclusion of HF exchange leads to stronger electron binding and mitigates the underbinding errors of GGA. It has been suggested that HF pseudopotentials may be useful in a variety of contexts, such as modeling systems with negatively-charged reference states~\cite{Al-Saidi08p075112} and in diffusion Monte Carlo simulations~\cite{Greeff98p1607,Ovcharenko01p7790}.
The successful development of HF pseudopotentials~\cite{Al-Saidi08p075112} has opened the possibility of constructing hybrid pseudopotentials by including an exact exchange component into GGA potentials.  Previous work demonstrated PBE0 pseudopotentials for gallium, indium and nitrogen atoms~\cite{Wu09p115201}. However, such potentials are  simple linear combinations of the exact exchange potential and the GGA derived potential without self-consistently solving hybrid PBE0 all-electron calculations.

In this paper, we construct self-consistent pseudopotentials (Sec.~\ref{sec_thy}) with the PBE0 hybrid density functionals, following the Rappe-Rabe-Kaxiras-Joannopoulos (RRKJ) method~\cite{ Rappe90p1227}.
We benchmark the hybrid functional pseudopotential accuracy for diatomic molecules in the G2 dataset and for simple solids, focusing on geometric parameters and fundamental gaps (Sec.~\ref{sec_test}). Consistent use of the density functional between pseudopotential and molecular/solid calculations generally reduces the error by 0.1$\%$ on bond lengths and 3$\%$ on HOMO-LUMO gaps.  
The PBE0 pseudopotential generator is implemented in the OPIUM software package~\cite{Opium}.


\section{Theoretical Methods}\label{sec_thy}
 
In this section, we provide an overview of the standard theory behind pseudopotential construction, before discussing the special considerations that must be taken into account for hybrid functional pseudopotentials.

\subsection{Pseudopotential construction}\label{sec_const}

The all-electron (AE) wavefunctions and eigenvalues of an atom are the foundation for the construction of all pseudopotentials.
The AE Kohn-Sham (KS) equation is
\begin{equation}
\left[-\frac{1}{2}\bigtriangledown^2+V_{\text{ion}}(\mathbf{r})+V_{\text{H}}[\rho(\mathbf{r})]+V_{\text{xc}}[\rho(\mathbf{r})]\right]\psi_i^{\text{AE}}(\mathbf{r})=\epsilon_i^{\text{AE}}\psi_i^{\text{AE}}(\mathbf{r}),
\label{eq:ae}
\end{equation} 
where $-\frac{1}{2}\bigtriangledown^2$ is the single-particle kinetic-energy operator, $V_{\text{ion}}(\mathbf{r})$ is the ionic potential that electrons feel from the nucleus, $V_{\text{H}}[\rho(\mathbf{r})]$ is the Hartree potential, and  $V_{\text{xc}}[\rho(\mathbf{r})]$ is the exchange-correlation potential,  which are functionals of the charge density $\rho(\mathbf{r})$. The all-electron wavefuction is denoted by $\psi_i^{\text{AE}}(\mathbf{r})$, and the all-electron energy eigenvalues by $\epsilon_i^{\text{AE}}$. For an atom, $V_{\text{ion}}(\mathbf{r})=-\frac{Z}{r}$, where $Z$ is the nuclear charge. Representing the wavefunction in spherical coordinates, 
$r=|\mathbf{r}|$ and each $\psi^{\text{AE}}_i(\mathbf{r})$ can be written as,
\begin{equation}
\psi^{\text{AE}}_{nlm}(\mathbf{r})=\frac{\phi^{\text{AE}}_{nl}(r)}{r}Y_{lm}(\theta,\phi),
\label{eq:spherical_coord}
\end{equation}
where $n,l,m$ are principal, angular, and spin quantum numbers, and $\theta$ and $\phi$ are the corresponding angles from spherical coordinates. $\phi^{\text{AE}}_{nl}$ is the radial wavefunction and $Y_{lm}(\theta,\phi)$ are the spherical harmonics. Now, Eq.~\ref{eq:ae} can be simplified in terms of $\phi_{nl}$: 
\begin{equation}
\left(-\frac{1}{2}\frac{d^2}{dr^2}+\frac{l(l+1)}{r^2}+V_{\text{KS}}(r)\right)\phi^{\text{AE}}_{nl}(r)=\epsilon^{\text{AE}}_{nl}\phi^{\text{AE}}_{nl}(r),
\label{eq:radial}
\end{equation}
where $V_{\text{KS}}(r)=V_{\text{ion}}(r)+V_{\text{H}}(r)+V_{\text{xc}}(r)$.
Instead of solving the full all-electron KS equation as in (Eq.~\ref{eq:ae}), it is computationally more efficient to solve the radial equation (Eq.~\ref{eq:radial}) self-consistently to obtain the radial wavefunction, $\phi^{\text{AE}}_{nl}(r)$ and corresponding eigenvalue, $\epsilon^{\text{AE}}_{nl}$.


In most molecular or solid systems, the valence electrons of atoms within the system are more crucial than core electrons, because they are more involved in chemical bonding.  
The core electrons mostly contribute to the electrostatic shielding of the nucleus. The AE wavefunctions of core electrons can contain rapid oscillations, which will cause difficulty in solving Eq.~\ref{eq:radial} numerically.
Therefore, it is advantageous to construct pseudopotentials, which capture the valence electron behavior and also eliminate the need to recalculate the core electron wavefunctions. 

Replacing the potential by a pseudopotential operator, the KS equation can be written as,
\begin{equation}
\left[-\frac{1}{2}\frac{d^2}{dr^2}+\frac{l(l+1)}{2r^2}+\hat{V}_{\text{PS}}\right]\phi^{\text{PS}}_{nl}(r)=\epsilon^{\text{PS}}_{nl}\phi^{\text{PS}}_{nl}(r),
\label{eq:radialps}
\end{equation}
where $\hat{V}_{\text{PS}}$ is the screened pseudopotential operator. Note that such an operator is usually non-local (is an integral operator on $\phi^{\text{PS}}_{nl}(r)$). Similar to $V_{\text{KS}}$, $\hat{V}_{\text{PS}}=\hat{V}^{\text{PS}}_{\text{ion}}+{V}_{\text{H}}(r)+{V}_{\text{xc}}(r)$. $\epsilon^{\text{PS}}_{nl}$ is the pseudo-eigenvalue,  and $\phi^{\text{PS}}_{nl}(r)$ is the pseudo-wavefunction. 
Norm-conserving pseudo-wavefunctions~\cite{Hamann79p1494}  should obey the following criteria:
\begin{equation*}
    \begin{aligned}
(1)\quad& \phi^{\text{PS}}_{nl}(r) = \phi^{\text{AE}}_{nl}(r), \quad
 \frac{d\phi^{\text{PS}}_{nl}(r)}{dr}=\frac{d\phi^{\text{AE}}_{nl}(r)}{dr}, \quad
 \frac{d^2\phi^{\text{PS}}_{nl}(r)}{dr^2}=\frac{d^2\phi^{\text{AE}}_{nl}(r)}{dr^2}\text{ for } r\geqslant r_c. \\
 (2)\quad& \epsilon^{\text{PS}}_{nl} =\epsilon^{\text{AE}}_{nl} \\
 (3)\quad& \langle\phi^{\text{PS}}_{nl}|\phi^{\text{PS}}_{nl}\rangle =\langle\phi^{\text{AE}}_{nl}|\phi^{\text{AE}}_{nl}\rangle =1 \\
 (4)\quad& \frac{d}{d\epsilon}\left(\frac{d\ln\phi^{\text{PS}}_{nl}(r)}{dr}\right)\bigg |_{R,\epsilon_{nl}} = \frac{d}{d\epsilon}\left(\frac{d\ln\phi^{\text{AE}}_{nl}(r)}{dr}\right)\bigg |_{R,\epsilon_{nl}},\ R\geqslant r_c
    \end{aligned}
\end{equation*}

Together, they guarantee wavefunction smoothness and continuity, that the solutions of the pseudo-system are accurate representations of the corresponding all-electron system, and that the error of eigenenergy shifts caused by chemical bonding is small for gentle changes to the wavefuntions and density~\cite{Hamann79p1494}, hence improving the transferability, or applicability of the pseudopotential in different chemical environments. 

In the RRKJ method~\cite{Rappe90p1227}, the pseudo-wavefunction is constructed as a sum of $N_b$ spherical Bessel functions $j_l(q_k r)$:
\begin{equation}
\phi^{\text{PS}}_{nl}(r)=
  \begin{cases}
    \sum^{N_b}_{k=1}c_{nlk}r j_{l}(q_k r),      & \quad r < r_c \\
   \phi^{\text{AE}}_{nl}(r), & \quad r\geqslant r_c\\
  \end{cases}
  \label{eq:RRJK}
\end{equation}
where the coefficients, $c_{nlk}$, are chosen to normalize the wavefunction and satisfy continuity constraints at $r_c$. 
Additional $c_{nlk}$ coefficients improve plane-wave convergence. Once the pseudo-wavefunction is constructed, the pseudopotential is obtained by inverting the pseudo-KS equation above (see Eq.(\ref{eq:radialps})). In applications of the pseudopotential in solid-state or molecular calculations, the screening effect of the valence electrons will generally be different from in the atomic calculation. Therefore, the valence electron screening is removed to obtain a descreened pseudopotential, $V^{\text{PS}}_{\text{ion},l}(r)$ for each angular momentum $l$, by subtracting Hartree and exchange-correlation potentials from the screened pseudopotential

\begin{equation}\label{eq:descreenps}
V^{\text{PS}}_{\text{ion},l}(r)= V^{\text{PS}}_l(r)-V_{\text{H}}[\rho_{\text{val}}](r)-V_{\text{xc}}[\rho_{\text{val}}](r),
\end{equation}
where $V_{\text{H}}[\rho_{\text{val}}](r)$ and $V_{\text{xc}}[\rho_{\text{val}}](r)$ are calculated only from the valence charge density. The full pseudopotential, written in semilocal form, is then

\begin{equation}
\begin{aligned}
\hat{V}^{\text{PS}}_{\text{ion}} =& \sum_{lm} V^{\text{PS}}_{\text{ion},l}(r)\, \lvert Y_{lm} \rangle \langle Y_{lm} \rvert \\
=& V_{\textrm{loc}}(r) + \sum_l\Delta \hat{V}_l^{\textrm{SL}} \\
\end{aligned}
\end{equation}

\noindent In the second line, the potential is expressed as the sum of a local potential $V_{\textrm{loc}}(r)$ and semilocal corrections $\Delta\hat{V}_l^{\textrm{SL}}$, which are projections in the angular coordinates yet local in the radial coordinate. In order to reduce the memory cost of computation, we write the semilocal pseudopotential in a fully-separable nonlocal Kleinman-Bylander~\cite{Kleinman82p1425} form 

\begin{equation}\label{eq:KB}
\begin{aligned}
\hat{V}^{\textrm{PS}} =& \hat{V}^{\textrm{loc}} + \sum_l \Delta \hat{V}_l^{\textrm{NL}} \\
\Delta \hat{V}_l^{\textrm{NL}} =& \frac{\Delta \hat{V}_l^{\textrm{SL}} \lvert \phi_{nl}^{\textrm{PS}} \rangle \langle \phi_{nl}^{\textrm{PS}} \rvert \Delta \hat{V}_l^{\textrm{SL}}}{\langle \phi_{nl}^{\textrm{PS}} \rvert \Delta \hat{V}_l^{\textrm{SL}} \lvert \phi_{nl}^{\textrm{PS}} \rangle} 
\end{aligned}
\end{equation}

\noindent
Writing the pseudopotential in this form ensures that semilocal and nonlocal pseudoatoms have the same eigenvalues and wavefunctions for the reference configuration. 
The transferability of such a nonlocal pseudopotential, to configurations other than the reference, can be improved by applying the designed nonlocal strategy, which involves modifying the projectors of Eq.~\ref{eq:KB}~\cite{Ramer99p12471}. 



\subsection{Hartree-Fock pseudopotentials}\label{sec:HF_ps}
Pseudopotentials can be constructed by solving the all-electron (AE) and pseudopotential (PSP) equations, Eq.~\ref{eq:ae} and Eq.~\ref{eq:radialps}
, above using different exchange-correlation functionals, such as LDA or GGA. It is crucial that the exchange-correlation functional used for pseudopotential construction is the same as the functional used in the target calculation~\cite{Fuchs98p2134}. When the exchange-correlation functional contains the Fock operator, as is the case for the hybrid functionals presently in widespread use, there are special considerations that must be taken into account in constructing the pseudopotential.Here, we consider the case of Hartree-Fock 
(HF) pseudopotentials, where the exchange-correlation functional is just the Fock operator, and will examine the PBE0 hybrid functional in the next subsection, where the Fock operator and PBE exchange-correlation are combined. For the HF pseudopotential, instead of solving the KS equation as in Eq.(\ref{eq:radial}), we solve the Hartree-Fock equation,

\begin{equation}
\left(\hat{T}+V_{\text{ion}}(\mathbf{r})+\hat{V}_{\text{HF}}[\{ \psi_{n'l'}\}]\right)\psi_{nl}(\mathbf{r})=\epsilon_{nl}\psi_{nl}(\mathbf{r}),
\label{eq:hf}
\end{equation}
where $\psi_{nl}(\mathbf{r})$ still takes the form in Eq.(\ref{eq:spherical_coord})  (dropping the AE superscript for simplicity), $V_{\text{ion}}(\mathbf{r})$ is the ionic potential, and $\hat{V}_{\text{HF}}[\{ \psi_{nl}\}]$ is the HF potential, which depends on the set of wavefunctions $\{ \psi_{nl}\}$. It is separated into two terms, 
\begin{equation}
\hat{V}_{\text{HF}}[\{ \psi_{n'l'}\}]=\hat{V}_{\text{H}}[\{ \psi_{n'l'}\}]+\hat{V}_{\text{x}}[\{ \psi_{n'l'}\}].
\end{equation}
The Hartree potential takes the form 
\begin{equation}
\langle \psi_{nl}|\hat{V}_{\text{H}}[\{ \psi_{n'l'}\}]|\psi_{nl}\rangle=\sum_{n'l'}\int d^3\mathbf{r}'d^3\mathbf{r}\frac{|\psi_{n'l'}(\mathbf{r}')|^2|\psi_{nl}(\mathbf{r})|^2}{|\mathbf{r}-\mathbf{r}'|},
\end{equation}
and the exact exchange operator acts as 

\begin{equation}
\langle \psi_{nl}|\hat{V}_{\text{x}}[\{ \psi_{n'l'}\}]|\psi_{nl}\rangle= \sum_{n'l'}\int d^3\mathbf{r}'d^3\mathbf{r}\frac{\psi_{nl}(\mathbf{r})\psi^{*}_{n'l'}(\mathbf{r})\psi_{n'l'}(\mathbf{r}')\psi^{*}_{nl}(\mathbf{r}')}{|\mathbf{r}-\mathbf{r}'|}.
\label{eq:vx} 
\end{equation}

Direct evaluation of the Fock integral above (Eq.~\ref{eq:vx}) requires introduction of angular variables for orbitals with non-zero angular momentum. This would result in non-spherical pseudopotentials, as well as introduce complexity into the pseudopotential generation process, which would then depend on the exact atomic configuration, including magnetic quantum numbers. To circumvent these issues, we make use of a spherical approximation, to construct spherical Hartree-Fock pseudopotentials. Spherical approximations are routinely used to construct spherical LDA and GGA pseudopotentials, which are widely used successfully in electronic and structural calculations.

We use the Hartree-Fock spherical approximation of Froese  Fischer~\cite{Fischer87p355} based on the concept of the ``average energy of configuration" introduced by Slater~\cite{Slater}. Consider all atomic configurations where the $i$-th shell, with principal and total angular quantum numbers $n_i$ and $l_i$, is occupied with weight $w_i$. That is, all permutations of $w_i$ electrons occupying the $(2l_i+1)$-degenerate shell $(n_il_i)$.

The average energy of all such atomic configurations, expressed as a sum over pairs of atomic orbitals $(n_il_i)$ and $(n_jl_j)$, is
\begin{equation}
\begin{aligned}
E_{\text{av}}^{\text{HF}}&=\sum_{i=1}^{m} w_i[I(n_il_i,n_il_i)+\left(\frac{w_i-1}{2}\right)\sum_{k=0}^{2l_i}f_k(l_i,l_i)F^{k}(n_il_i,n_il_i)]\\
&+\sum_{i=2}^{m}\left\{\sum_{j=1}^{i-1}w_iw_j\left[F^0(n_il_i,n_jl_j)+\sum_{k=|l_i-l_j|}^{(l_i+l_j)}g_k(l_i,l_j)G^k(n_il_i,n_jl_j)\right]\right\},
\label{eq:slater_average}
\end{aligned}
\end{equation}
Here, the first summation  represents the one electron contribution, 

\begin{equation}
I(nl,nl)=-\frac{1}{2}\int_o^{\infty}\phi_{nl}^{*}(r)\left(\frac{d^2}{dr^2}+\frac{2Z}{r}-\frac{l(l+1)}{r^2}\right)\phi_{nl}(r)dr.
\label{eq:slater_oneel}
\end{equation}

The other terms contain the interaction terms between pairs of electrons. $F^k$ and $G^k$ are the Hartree and exchange energy Slater integrals,

\begin{equation}\label{eq:slater_Fk}
F^{k}(nl;n'l')=\int_{0}^{\infty}\int_{0}^{\infty}\phi_{nl}(r)\phi_{nl}(r)\frac{r_<^k}{r_>^{k+1}}\phi_{n'l'}(r')\phi_{n'l'}(r')drdr',
\end{equation}
and
\begin{equation}
G^{k}(nl;n'l')=\int_{0}^{\infty}\int_{0}^{\infty}\phi_{nl}(r)\phi_{n'l'}(r')\frac{r_<^k}{r_>^{k+1}}\phi_{n'l'}(r)\phi_{nl}(r')drdr',
\label{eq:slater_FandG}
\end{equation}


\noindent
where $r_<$ ($r_>$) is the lesser (greater) of $r$ and $r'$. Details of the derivation are provided in Appendix~C, and the numerical coefficients $f_k$ and $g_k$ are tabulated in Ref.~\cite{Slater}. We note that the integrals in Eq.~\ref{eq:slater_oneel}--\ref{eq:slater_FandG} for the average energy depend only on the radial coordinate, and hence are a simplification of Eq.~\ref{eq:vx}.  

Taking functional derivatives of Eq.~\ref{eq:slater_average} with respect to the radial wavefunctions $\phi_i(r)$, we arrive at Hartree-Fock equations for the wavefunctions of a Hartree-Fock atom.
The set of $m$ radial wavefunctions $\phi_{i}, \, i=1,\dots,m$ obeys the coupled set of equations 

\begin{equation}
\hat{L}\,\phi_{i}(r)=\frac{2}{r} \, \big [ Y_i[\{\phi\}](r) \, \phi_{i}(r)+X_i[\{\phi\}](r) \big ]+\sum_{j=1}^{m} \varepsilon_{ij}\phi_{j}(r),
\label{eq:coupled_hf}
\end{equation}
where $\hat{L}=\frac{d^2}{dr^2}-2V_{\text{ion}}(r)-\frac{l_i(l_i+1)}{r^2}$ is the single-particle part of the Hartree-Fock Hamiltonian, $(2/r)Y_i[\{\phi\}](r)$ and $(2/r)X_i[\{\phi\}](r)$ are the Hartree and exchange terms~\cite{Fischer}, $\varepsilon_{ij}$ are Lagrange multipliers for orthogonality and normalization of radial wavefunctions. The detailed derivation of all these terms are presented in Appendix D.  

Once the HF equation is constructed, we solve these equations self-consistently in a similar way to DFT pseudopotentials. The HF pseudowavefunctions $\phi_{nl}^{\text{PS}}(r)$ are constructed using the same RRKJ procedure (Eq.(\ref{eq:RRJK})) as for the DFT pseudowavefunctions. The screened pseudopotential is obtained by inverting Eq.(\ref{eq:hf}). Similar to DFT pseudopotentials, we descreen by subtracting the Hartree and exchange contributions of the valence electrons (c.f. Eq.~\ref{eq:descreenps})  
\begin{equation}
V_{\text{ion},l}^{\text{PS}}(r)
=V_{l}^{\text{PS}}(r)-
\frac{2}{r} Y_i[\{\phi_{\text{val}}\}](r)-
\frac{2X_i[\{\phi_{\text{val}}\}](r)}{r\phi_{i}(r)} , 
\end{equation}
with $Y_i$ and $X_i$ obtained from Eq.~\ref{eq:coupled_hf}.
The HF pseudopotential constructed this way has a long-range non-Coulombic component of the tail, which does not decay as $1/r$. This is a consequence of the non-local nature of the Fock operator~\cite{Al-Saidi08p075112}. To resolve this issue, we make use of the localization procedure of Trail and Needs~\cite{Trail05p014112}. The tail is forced to asymptotically approach $1/r$, and the potential is modified within the localization radius to ensure consistency with the all-electron eigenvalues~\cite{Al-Saidi08p075112}.

\subsection{PBE0 pseudopotentials}
As hybrid functionals are a mix of HF and DFT ingredients, we generate hybrid pseudopotential using the HF pseudopotential approach as a foundation. 
The PBE0 density functional~\cite{Perdew96p9982} was developed based on the PBE exchange-correlation functional~\cite{Perdew96p3865}; the PBE0 form is

\begin{equation}\label{eq:pbe0}
E_{\text{xc}}^{\text{PBE0}}=aE_{\text{x}}^{\text{HF}}+(1-a)E_{\text{x}}^{\text{PBE}}+E_{\text{c}}^{\text{PBE}},
\end{equation}
where $a=0.25$ for the PBE0 functional. As we use the spherical approximation for $E_{\text{x}}^{\text{HF}}$ (Eq.~\ref{eq:slater_average}), we likewise evaluate the PBE exchange-correlation functional using a spherical approximation. Since $E_\text{x}^{\text{PBE}}$ is a functional of density only, this method consists of evaluating $E_\text{x}^{\text{PBE}}$ in Eq.~\ref{eq:pbe0} at the charge density, again taken to be the average over all possible magnetic quantum number configurations.

\begin{equation}
\rho_{nl}(r)= \sum_{nlm} f_{nlm} |\psi(\mathbf{r})_{nl}|^2  = \frac{1}{4\pi}\sum_{n_il_i} f_{n_il_i}|\phi_{n_il_i}(r)|^2,
\end{equation}
where $\rho_{nl}(r)$ is the spherical symmetric charge density, $f_{n_il_i}=w_i$ (as in Appendix~B) is the occupation number for each orbital $(n_il_i)$, and $f_{nlm}=f_{nlm'}$ is the occupation number for each magnetic quantum number $(nlm)$. Upon including $E_{\text{x}}^{\text{PBE}}$ and $E_{\text{c}}^{\text{PBE}}$ into the total energy expression Eq.~\ref{eq:slater_average},
and taking functional derivatives, the coupled set of HF equations (Eq.~\ref{eq:coupled_hf}) becomes

\begin{equation}
\hat{L}\phi_{i}(r)=\frac{2}{r}[Y_i(r)\phi_{i}(r)+\frac{1}{4}X_i(r)]+\frac{3}{4}V_{\text{x}}^{\text{PBE}}(r)+V_{\text{c}}^{\text{PBE}}(r)+\sum_{j=1}^{m}\delta_{l_il_j}\epsilon_{ij}\phi_{j}(r),
\end{equation}
where the additional terms are the PBE exchange potential $V_{\text{x}}^{\text{PBE}}(r)$ and the  PBE correlation potential $V_{c}(r)$. 
The self-consistent solution of these coupled equations is found iteratively, in a similar fashion to the HF equations (Eq.~\ref{eq:coupled_hf}). At each iteration, we calculate the Fock exchange term ($X_i(r)$) from the wavefunctions of the previous iteration, and the PBE terms ($V_\text{x}^{\text{PBE}}$, $V_\text{c}^{\text{PBE}}$) from the density of the previous iteration. The pseudopotential construction is performed the same way as for HF pseudopotentials, including RRKJ pseudization, descreening, and localization of the non-Coulombic tail.


%


\section{Testing of PBE0 pseudopotentials on molecular and solid state systems}\label{sec_test}

We test the accuracy of our PBE0 pseudopotentials and the importance of pseudopotential density functional consistency for PBE0. We compare PBE calculations using PBE pseudopotentials (PBE), PBE0 calculations using PBE0 pseudopotentials (PBE0) and PBE0 calculations using PBE pseudopotentials (PBE-PBE0). The last case is currently the most widely used method of performing PBE0 calculations. The DFT code we use is Q\textsc{uantum-espresso}~\cite{Giannozzi09p395502}. Each single molecule is put into 20.0 \AA\ cubic box, and its energy and geometry computed with kinetic energy cutoff $E_{\text{cut}}$=25.0 Hartree. All these calculations are spin-polarized. The total energy convergence and force convergence are set to 0.005 mHartree/cell and 0.05 mHartree/\AA. The reference all-electron calculations are performed using FHI-aims~\cite{Blum09p2175} with tight basis settings. The molecular and crystal structural optimizations are converged within 3 $\times10^{-3}$ mHartree/cell for total energy, and the forces are converged within 0.003 mHartree/\AA. 

In Table \ref{tab:2}, we show the bond lengths for diatomic molecules that belong to G2 data set~\cite{Adamo99p6158} and compare each of our pseudopotential calculations with PBE0 all-electron values~\cite{Becke98p2092}. 
The PBE functional gives the worst mean absolute relative error (MARE) 
of 1.08$\%$ when comparing to FHI-aims PBE0. 
The use of PBE pseudopotential in PBE0 calculation gives MARE of 0.71$\%$. Using the PBE0 functional with the PBE0 pseudopotential, the MARE reduces to $0.53\%$. 
This indicates that pseudopotential density functional consistently improves bond lengths for PBE0. 
\begin{table}[h]
\centering
\caption{The bond lengths of the diatomic molecules from G2 data set calculated from PBE, PBE-PBE0 and PBE0. The all-electron data are calculated using FHI-aims~\cite{Blum09p2175}. Units in \AA. The MARE is calculated as MARE$=\frac{1}{N}\sum_i^N\frac{|b_i-b_{\text{AE}}|}{b_{\text{AE}}}\times100$, where $N$ is the number of species, $b_i$ is the bond length of each species, and $b_{\text{AE}}$ is the PBE0 all-electron value.
}
\label{tab:2}
\begin{tabular}{lccccc}
\hline
\hline
Molecule & PBE & PBE-PBE0 & PBE0 & AE-PBE & AE-PBE0 \\
\hline
H$_2$   & 0.753       & 0.747       & 0.747       & 0.750        & 0.746 \\
LiH  & 1.600         & 1.595       & 1.596       & 1.603       & 1.595 \\
BeH  & 1.348       & 1.343       & 1.351       & 1.355       & 1.348 \\
CH   & 1.137       & 1.122       & 1.122       & 1.136       & 1.124 \\
NH   & 1.070        & 1.056       & 1.041       & 1.050        & 1.041 \\
OH   & 0.983       & 0.975        & 0.966       & 0.983       & 0.983 \\
FH   & 0.928       & 0.914       & 0.912       & 0.93        & 0.918 \\
Li$_2$  & 2.719       & 2.725       & 2.718       & 2.728       & 2.723 \\
LiF  & 1.578       & 1.567       & 1.566       & 1.574       & 1.562 \\
CN   & 1.174       & 1.159       & 1.159       & 1.175       & 1.159 \\
CO   & 1.135       & 1.123       & 1.122       & 1.136       & 1.122 \\
N$_2$   & 1.081       & 1.069       & 1.069       & 1.103       & 1.089 \\
NO   & 1.132       & 1.113       & 1.138       & 1.157       & 1.139 \\
O$_2$   & 1.212       & 1.218       & 1.217       & 1.218       & 1.192 \\
F$_2$   & 1.420        & 1.382       & 1.382       & 1.413       & 1.376 \\
\hline
MARE & 1.08 & 0.71 & 0.53 & 1.08 & \\
\hline
\hline     
\end{tabular}
\end{table}

One of the reasons for using hybrid density functionals is that they predict fundamental gaps and ionization potentials (IP) more accurately than the PBE functional \cite{Ernzerhof99p5029,Matsushita11p075205,Wu09p115201}. 
Table \ref{tab:comp_IP} shows the HOMO eigenvalues for diatomic molecules within the G2 dataset, calculated from different density functionals and compared with HOMO levels calculated from all-electron calculations. The MARE between PBE HOMO eigenvalues and all-electron PBE0 values is the largest among the three computed cases. Both PBE0 cases are smaller than PBE case, and the MARE of PBE0 is reduces by 0.13$\%$ compare to PBE-PBE0.
\begin{table}[h]
\centering
\caption{HOMO eigenvalues with PBE, PBE-PBE0 and PBE0 methods. Energies are in eV. 
The all-electron PBE0 values are used as the reference.}
\label{tab:comp_IP}
\begin{tabular}{lccccc}
\hline
\hline
Molecule & PBE & PBE-PBE0 & PBE0 & AE-PBE & AE-PBE0 \\
\hline
H$_2$       & -10.31 & -11.96     & -11.96  & -10.34       & -11.99        \\
LiH      & -3.89  & -5.45      & -5.44   & -4.35        & -5.44         \\
BeH      & -4.76  & -5.77      & -5.20   & -4.68        & -5.69         \\
CH       & -5.91  & -7.43      & -7.43   & -5.84        & -7.45         \\
NH       & -7.98  & -9.78      & -9.76   & -6.69        & -9.76        \\
OH       & -7.06  & -8.81      & -8.72   & -7.14        & -7.00         \\
FH       & -9.33  & -11.43     & -11.43  & -9.61        & -11.86        \\
Li$_2$      & -3.20  & -3.99      & -3.75   & -3.16        & -3.72         \\
LiF      & -6.08  & -7.77      & -7.85   & -6.09        & -7.96         \\
CN       & -9.30  & -10.74     & -10.94  & -9.38        & -9.32         \\
CO       & -9.01  & -10.41     & -10.42  & -9.03        & -10.72        \\
N$_2$       & -10.07 & -11.93     & -12.20  & -10.22       & -12.20        \\
NO       & -4.74  & -6.25      & -6.29   & -4.50        & -4.60         \\
O$_2$       & -6.71  & -8.68      & -8.70   & -6.91        & -8.91         \\
F$_2$       & -9.41  & -11.50     & -11.58  & -9.46        & -11.68        \\
\hline
MARE     & 15.87   & 6.79       & 6.66    &     16.06         & \\
\hline
\hline

\end{tabular}
\end{table}
In Table \ref{tab:comp_HL}, we present the HOMO-LUMO gap for the same dataset as in Table \ref{tab:comp_IP}. 
\begin{table}[h]
\centering
\caption{HOMO-LUMO gap (in eV) of diatomic molecules in G2 dataset with different functionals. The PBE0 all-electron results are used as the reference.
}
\label{tab:comp_HL}
\begin{tabular}{lccccc}
\hline
\hline
Molecule & PBE & PBE-PBE0 & PBE0 & AE-PBE & AE-PBE0 \\
\hline
H$_2$   & 10.26  & 11.94 & 11.94 & 10.84 & 13.10 \\
LiH  & 2.57   & 4.04  & 4.48  & 2.81  & 4.45  \\
BeH  & 2.64   & 4.44  & 4.42  & 2.31  & 4.15  \\
CH   & 2.06   & 3.95  & 3.51  & 1.77  & 3.60  \\
NH   & 3.95   & 7.27  & 7.34  & 6.45  & 7.16  \\
OH   & 1.12   & 4.77  & 4.92  & 6.54  & 4.25  \\
FH   & 8.19   & 10.92 & 10.93 & 8.76  & 11.80 \\
Li$_2$  & 1.41   & 2.75  & 2.47  & 1.43  & 2.50  \\
LiF  & 4.29   & 6.41  & 6.50  & 4.62  & 7.02  \\
CN   & 1.99   & 4.67  & 4.74  & 1.72  & 4.48  \\
CO   & 6.98   & 9.61  & 9.62  & 6.98  & 10.04 \\
N$_2$   & 7.66   & 10.94 & 10.94 & 8.24  & 11.71 \\
NO   & 1.30   & 3.50  & 2.88  & 1.22  & 2.86  \\
O$_2$   & 2.40   & 5.74  & 6.09  & 2.31  & 6.10  \\
F$_2$   & 3.32   & 7.77  & 7.79  & 3.63  & 8.34  \\
\hline
MARE & 44.70 & 7.96 & 4.55 & 40.88 & \\
\hline
\hline
\end{tabular}
\end{table}
Both PBE0 cases gave much closer values to the AE PBE0 reference, and our PBE0 pseudopotential showed a small error reduction compared to the hybrid DFT calculated with PBE pseudopotentials. Similar to bond length calculations, the consistency of the exchange-correlation density functional between pseudoptential and DFT calculation reduces the error. This indicates that the use of PBE pseudopotential for PBE0 DFT calculation results in good accuracy, which can be improved by implementing the corresponding pseudopotential with a consistent density functional. 

We have also tested our pseudopotentials in solid-state calculations. The lattice constants and band gaps for $\alpha$-Si and $\beta$-GaN are shown in Table \ref{tab:4}. 
Similar to molecular bond lengths, the density functional consistency also influences the lattice constants of solids.
The lattice constant of $\alpha$-Si is slightly improved by using PBE0 pseudopotentials instead of PBE-PBE0.  
The PBE calculation significantly underestimates the band gaps. The two PBE0 cases increase the band gaps by a large amount compared to PBE calculation. The band gaps from PBE-PBE0 and PBE0 are within 1$\%$ of each other, for both Si and GaN. The PBE0 pseudopotential band gap tends to be lower, and closer to the experimental value. Together with the calculations from molecular properties, we may conclude that the systematic error from pseudopotential density functional inconsistency is of the order of 1$\%$ for PBE0, for the systems tested.  

\begin{table}[h]
\centering
\caption{Solid state calculation with PBE, PBE-PBE0 and PBE0. The lattice constant and band gap of Si and GaN are listed. The lattice constant is in units of \AA, and the band gap is in eV. The experimental band gaps are at 0K. Relative errors ($\%$) are listed in parentheses. All-electron PBE0 results are used as the reference.}


\label{tab:4}
\begin{tabular}{lccccc}
\hline
\hline
Crystal&	PBE&	PBE-PBE0&	PBE0 &AE-PBE& AE-PBE0 \\
\hline
Lattice constants&&&&\\
\hline
Si&	    5.484(0.219)&	5.452(0.073)&	5.446(-0.037)&	5.472(0.441) &5.448\\
GaN&	4.541(-0.176)&  4.539(0.066)&  4.537(0.022)&  4.549 (0.287)&4.536\\
\hline
Band Gap&&&&\\
\hline
Si&	0.58(-77.17)&	1.79(9.82)  &1.78(9.20)  &	2.54 (55.83)& 1.63\\
GaN&	1.81(16.77)&	3.58(1.13) &3.56(0.56)	 & 1.55 (-56.21)& 3.54\\     
\hline
\hline
\end{tabular}
\end{table}


%
\section{Conclusion}\label{sec_conc}
We have developed the first self-consistent PBE0 pseudopotential and have successfully implemented it in the OPIUM pseudopotential generation code. We have also shown that our PBE0 pseudopotentials behave well when implementing them to DFT calculations. 
Our benchmarking tests on G2 dataset indicate that the systematic error associated with pseudopotential density functional consistency is within 1$\%$. 
We have shown that using the PBE0 pseudopotential in PBE0 DFT calculations lead to improvements in bond length accuracy of 0.1$\%$ compared to PBE0 all-electron DFT calculations with PBE pseudopotentials. The HOMO eigenvalues for G2 dataset predicted by using PBE0 pseudopotentials are closer to the all-electron values compared to PBE-PBE0. On average, for our test set,
the error of HOMO-LUMO gaps for molecules is reduced by about 3$\%$. A similar trend is obtained for the solids tested. From these results, we conclude that using PBE pseudopotentials in PBE0 calculations leads to acceptable results for small molecules and simple solids, while using PBE0 pseudopotentials instead will likely result in a small consistent increase in accuracy. 
Future directions include further testing of PBE0 pseudopotentials for more complex systems, the inclusion of relativistic effects for heavy atoms, and the development of other hybrid functional pseudopotentials, including range-separated hybrids~\cite{Toulouse10p032502}.

\section{Acknowledgements}
J.Y. was supported by the U.S. National Science Foundation, under grant CMMI-1334241.
L.Z.T. was supported by the U.S. ONR under Grant N00014-17-1-2574.
A.M.R. was supported by the U.S. Department of Energy, under grant DE-FG02-07ER46431.
Computational support was provided by the HPCMO of the U.S. DOD and the NERSC of the U.S. DOE.

\section{Appendix A: Construction of PBE0 pseudopotentials on a real space grid}\label{sec_bench}

The accuracy of the real space pseudopotential generator depends on the radial grid size.
The use of the logarithmic grid ensures enough grid points near the core to describe oscillations of the all-electron wavefunctions in that region, while capturing the tail of the wavefunctions at large distances from the core to sufficient accuracy.   
The logarithmic grid is defined as 
\begin{equation}
r_i=aZ^{-1/3}e^{(i-1)b}, i=1,...,N
\end{equation}
where $N$ is the number of grid points, spanning a sufficiently large real space range ($r_{max}$), $Z$ is the core charge, and $a$ controls the position of the first grid point, and $b$ determines the grid spacing. We use values of $a=0.0001$ and $b=0.013$. The number of grid points $N$ is obtained by setting $r_{\rm{max}}$=80 Bohr. 

\section{Appendix B: Derivation of Hartree-Fock average energy}\label{sec:hfaverage}

As a preliminary to deriving the average energy formula Eq.~\ref{eq:slater_average}, we collect several useful quantities. The Hartree potential due to an electron in the state $(nlm)$ is

\begin{equation}\label{eq:vhnlm}
\begin{aligned}
V_H^{(nlm)}(\vec{r}) =& \int d^3r' \frac{|\psi_{nlm}(\vec{r'})|^2}{\lvert \vec{r} - \vec{r'}\rvert} 
=& \int_0^\infty r'^2dr' d\Omega' \frac{\phi_{nl}(r')^2 \lvert Y_{lm}(\Omega)\rvert ^2 }{\lvert \vec{r} - \vec{r'}\rvert} 
\end{aligned}
\end{equation}

\noindent Using the expansion with $m$ here for getting ready for Eq.~\ref{eq:vhnlm-simple}

\begin{equation}\label{eq:expand}
\frac{1}{\lvert \vec{r} - \vec{r'}\rvert} = \sum_{k=0}^\infty \sum_{m=-k}^{k} \frac{4\pi}{2k+1} (-1)^m 
\frac{r_<^k}{r_>^{k+1}} Y_k^{-m}(\Omega) Y_k^{m}(\Omega')
\end{equation}

\noindent where $r_<$ ($r_>$) is the lesser (greater) of $r$ and $r'$, we write Eq.~\ref{eq:vhnlm} as

\begin{equation}\label{eq:vhnlm-simple}
\begin{aligned}
V_H^{(nlm)}(\vec{r}) =& \sum_{km'}  \int_0^\infty r'^2dr'
\frac{r_<^k}{r_>^{k+1}}
\sqrt{\frac{4\pi}{2k+1}}\, Y_k^{0*}(\Omega) \,
c^k(l,m',l,m') \, \phi_{nl}(r')^2 \\
=& \int_0^\infty r'^2dr' \frac{1}{r_>} \phi_{nl}(r')^2  
+ \sum_{k=1}^{2l} \sum_{m'}  \int_0^\infty r'^2dr'
\frac{r_<^k}{r_>^{k+1}}
\sqrt{\frac{4\pi}{2k+1}}\, Y_k^{0*}(\Omega) \,
c^k(l,m',l,m') \, \phi_{nl}(r')^2 
\end{aligned}
\end{equation}

\noindent Here, we make use of the symbols 

\begin{equation}\label{eq:gaunt}
\begin{aligned}
c^k(l,m,l',m') =& \sqrt{\frac{4\pi}{4k+1}} \int Y_{lm}^*(\Omega) Y_{k, m-m'}(\Omega) Y_{l'm'}(\Omega) d\Omega \\
=& (-1)^{-m} \sqrt{2l+1}\sqrt{2l'+1} \begin{pmatrix}l & k & l' \\ 0 & 0 & 0\end{pmatrix} \begin{pmatrix}l & k & l' \\ -m & m-m' & m'\end{pmatrix}
\end{aligned}
\end{equation}

\noindent for Gaunt's formula, in terms of Wigner $3j$-symbols. In the second line of Eq.~\ref{eq:vhnlm-simple}, we have separated the $k=0$ and $k>0$ components, because the latter vanishes when averaged over $m$. Therefore, the Hartree energy of a pair of electrons $(i j\vert i j)$, in orbitals $(n_i,l_i)$ and $(n_j,l_j)$, averaged over the magnetic quantum number $m_j$ of the second electron, is simply 

\begin{equation}\label{eq:avh}
\begin{aligned}
\langle (i j\vert i j) \rangle_{m_j} =& \int_0^\infty dr \phi_{n_i l_i}(r)^2 \int_0^\infty dr' \frac{1}{r_>} \phi_{n_j l_j}(r')^2 \\
=& F^0(n_i l_i,n_j l_j)
\end{aligned}
\end{equation}

\noindent The exchange integral for a pair of electrons in orbitals $(n_i,l_i)$ and $(n_j,l_j)$ can be calculated in similar fashion. Using Eqs.~\ref{eq:expand} and \ref{eq:gaunt}, we get

\begin{equation}\label{eq:exchint}
\begin{aligned}
(i j\vert j i) =& \int d^3r d^3r' \frac{\psi_{n_i l_i m_i}^*(\vec{r})\psi_{n_j l_j m_j}(\vec{r}) \psi_{n_j l_j m_j}^*(\vec{r'}) \psi_{n_i l_i m_i}(\vec{r'}) }{\lvert \vec{r} - \vec{r'}\rvert} \\
=&  \sum_{kq} \int Y_{l_i m_i}^*(\Omega) Y_{l_jm_j}(\Omega) Y_{kq}(\Omega) d\Omega
\int Y_{l_jm_j}^*(\Omega') Y_{l_im_i}(\Omega') Y_{kq^*}(\Omega') d\Omega' \\ 
&\int \frac{r_<^k}{r_>^{k+1}} \frac{4\pi}{2k+1} 
\phi_{n_il_i}(r) \phi_{n_jl_j}(r) \phi_{n_jl_j}(r') \phi_{n_il_i}(r') dr dr' \\
=& \sum_{k} c^k(l_i,m_i,l_j,m_j)^2
\int \frac{r_<^k}{r_>^{k+1}} 
\phi_{n_il_i}(r) \phi_{n_jl_j}(r) \phi_{n_jl_j}(r') \phi_{n_il_i}(r') dr dr'
\end{aligned}
\end{equation}


\noindent For the average of the exchange integral over $m_j$, we get 

\begin{equation}\label{eq:avx}
\langle (i j\vert j i) \rangle_{m_j} = \frac{1}{\sqrt{(2l_i+1)(2l_j+1)}} \sum_{k} c^k(l_i,0,l_j,0) G^k(n_il_i,n_jl_j)
\end{equation}

To calculate the average total energy of an atomic configuration, we must consider the Hartree and exchange energies of all pairs of electrons. First consider the case where the electrons are in the same orbital ($n_i=n_j$, $l_i=l_j$). In this case, since $G^k(n_il_i,n_il_i)=F^k(n_il_i,n_il_i)$, we can combine Eqs.~\ref{eq:avh},~\ref{eq:slater_Fk} and ~\ref{eq:avx} to obtain

\begin{equation}\label{eq:same}
\langle (i j\vert i j) - (i j\vert j i) \rangle = \frac{w_i(w_i-1)}{2} \sum_k f_k(l_i,l_i) F^k(n_il_i,n_il_i)
\end{equation}

\noindent where the numerical coefficients $f_k(l_i,l_i)$ are obtained from those in  Eqs.~\ref{eq:avh},~\ref{eq:avx}, and the prefactor $\frac{w_i(w_i-1)}{2}$ is the number of different electron pairs in orbital $i$. 

For the case where the electrons in the pair are in different orbitals, the sum of Eqs.~\ref{eq:avh},~\ref{eq:avx} gives 

\begin{equation}\label{eq:diff}
\langle (i j\vert i j) - (i j\vert j i) \rangle = w_i w_j \left ( F^0(n_il_i,n_jl_j) + \sum_k g_k(l_i,l_j) G^k(n_il_i,n_jl_j) \right)
\end{equation}

\noindent where the coefficients $g_k(l_i,l_j)$ are given by Eq.~\ref{eq:avx}. Collecting the terms in Eqs.~\ref{eq:same},~\ref{eq:diff} with the single-particle energies results in the expression for the average total energy Eq.~\ref{eq:slater_average}

\section{Appendix C: Derivation of self-consistent Hartree-Fock equations}\label{sec:hfeqns}

If the orbitals are not necessarily normalized, the average energy (as defined in Sec.~\ref{sec:HF_ps}) derived in Sec.~\ref{sec:hfaverage} may be written in the form 

\begin{equation}\label{eq:hfgen}
E_{\textrm{av}}^{\textrm{HF}} = \sum_i \frac{w_i I(n_il_i,n_il_i)}{\langle n_il_i \vert n_il_i \rangle }
+ \sum_{i ; k} \frac{a_{iik} F^k(n_il_i,n_il_i)}{\langle n_il_i \vert n_il_i \rangle \langle n_il_i \vert n_il_i \rangle}
+ \sum_{i> j ; k} \frac{a_{ijk} F^k(n_il_i,n_jl_j)}{\langle n_il_i \vert n_il_i \rangle \langle n_jl_j \vert n_jl_j \rangle}
+ \sum_{i> j ; k} \frac{b_{ijk} G^k(n_il_i,n_jl_j)}{\langle n_il_i \vert n_il_i \rangle \langle n_jl_j \vert n_jl_j \rangle}
\end{equation}

\noindent We wish to find wavefunctions that minimize $E_{\textrm{av}}^{\textrm{HF}}$, under the constraint of wavefunction orthogonality. In other words, a pair of radial functions from orbitals with the same angular momentum, $(n_i,l_i)$ and  $(n_j,l_j)$ with $l_i=l_j$, must be orthogonal. Using the Lagrange multipliers $\lambda_{ij}$, we therefore search for the stationary solutions of the functional

\begin{equation}\label{eq:kfunc}
K = E_{\textrm{av}}^{\textrm{HF}}  + \sum_{i>j} \delta_{l_il_j} \lambda_{ij} \frac{\langle n_il_i \vert n_jl_j \rangle}{\langle n_il_i \vert n_il_i \rangle^{1/2} \langle n_il_i \vert n_il_i \rangle^{1/2}}
\end{equation} 

\noindent We now proceed to take functional derivatives of Eqs.~\ref{eq:hfgen}, \ref{eq:kfunc} with respect to variations in a radial function $\phi_{nl}(r)$. We note that only a subset of terms in Eq.~\ref{eq:hfgen} involve $nl$, and those that do all contain a factor of $\langle n_il_i \vert n_il_i \rangle^{-1}$. We can therefore write those terms in the form $\tilde{E}(nl) = \langle n_il_i \vert n_il_i \rangle^{-1} \tilde{F}(nl)$ with the variation

\begin{equation}\label{eq:delebar}
\delta\tilde{E}(nl) = \langle n_il_i \vert n_il_i \rangle^{-1} \delta \tilde{F}(nl)+ \delta [\langle n_il_i \vert n_il_i \rangle^{-1}] \tilde{F}(nl)
\end{equation}

\noindent and

\begin{equation}
\begin{aligned}
\delta \tilde{F}(nl) =& w_{nl} \delta I(nl) 
+ \sum_{k} a_{nl,nl,k} F^k(nl,nl) \delta [\langle nl \vert nl \rangle^{-1} ]
+ \sum_k \frac{a_{nl,nl,k} \delta F^k(nl,nl)}{\langle nl \vert nl \rangle}\\
&+ \sum_{n'l'\neq nl ; k} \frac{a_{nl,n'l',k} \delta F^k(nl,n'l')}{\langle n'l' \vert n'l' \rangle}
+ \sum_{n'l'\neq nl ; k} \frac{b_{nl,n'l',k} \delta G^k(nl,n'l')}{\langle n'l' \vert n'l' \rangle}
\end{aligned}
\end{equation}

\noindent Furthermore, we have 

\begin{equation}
\delta [\langle n_il_i \vert n_il_i \rangle^{-1}] = -2\int dr\, \frac{\phi_{nl}(r) \delta \phi_{nl}(r)}{\langle nl \vert nl \rangle^2}
\end{equation}

\noindent and

\begin{equation}
\delta F^k(nl,n'l') = 2(1+\delta_{nl,n'l'}) \int dr\, \phi_{nl}(r)\, \delta \phi_{nl}(r)\, \frac{1}{r}\, Y^k(n'l',nl,r)
\end{equation}

\begin{equation}
\delta G^k(nl,n'l') = 2 \int dr\, \phi_{n'l'}(r)\, \delta \phi_{nl}(r)\, \frac{1}{r}\, Y^k(nl,n'l',r)
\end{equation}

\noindent where

\begin{equation}
Y^k(nl,n'l',r) = \int_0^r ds\, \frac{s^k}{r^k} \, \phi_{nl}(s)\, \phi_{n'l'}(s) +\int_r^\infty ds\, \frac{r^{k+1}}{s^{k+1}}\,  \phi_{nl}(s)\, \phi_{n'l'}(s)
\end{equation}

\noindent Finally, the variation of the terms involving the Lagrange multipliers in Eq.~\ref{eq:kfunc} is

\begin{equation}\label{eq:dellam}
\delta \left [ \sum_{n'}  \lambda_{nl,n'l'} \frac{\langle nl \vert n'l \rangle}{\langle nl \vert nl \rangle^{1/2} \langle n'l \vert n'l \rangle^{1/2}} \right ]
= \sum_{n'}  \lambda_{nl,n'l'} \frac{\int dr\, \phi_{n'l}(r)\, \delta \phi_{nl}(r) }{\langle nl \vert nl \rangle^{1/2} \langle n'l \vert n'l \rangle^{1/2}}
\end{equation}

The variational principle requires that the variation $\delta K$ be stationary with respect to $\delta \phi_{nl}(r)$. Collecting Eqs.~\ref{eq:delebar}--\ref{eq:dellam}, we obtain the Hartree-Fock equations (Eq.~\ref{eq:coupled_hf}) where

\begin{equation}
Y_i(r) = \sum_{j,k} \frac{(1+\delta_{n_il_i,n_jl_j}) a_{n_il_i,n_jl_j,k} Y^k(n_jl_j,n_jl_j,r)}{w_i \langle n_jl_j \vert n_jl_j \rangle}
\end{equation}

\begin{equation}
X_i(r) = \sum_{j\neq i,k} \frac{ b_{n_il_i,n_jl_j,k} Y^k(n_il_i,n_jl_j,r) \phi_{n_jl_j}(r)}{w_i \langle n_jl_j \vert n_jl_j \rangle}
\end{equation}

\noindent and

\begin{equation}
\varepsilon_{ii} = \frac{2}{w_i} \left [ \tilde{E}(n_il_i)-\sum_k \frac{a_{n_il_i, n_il_i,k} F^k(n_il_i,n_il_i)}{\langle n_il_i \vert n_il_i \rangle^2} \right ]
\end{equation}

\begin{equation}
\varepsilon_{ij} = \frac{\lambda_{n_il_i,n_jl_j} \langle n_il_i \vert n_il_i \rangle^{1/2}}{w_i \langle n_jl_j \vert n_jl_j \rangle^{1/2} }
\end{equation}

\bibliography{rappecites}

\begin{thebibliography}{29}%
\makeatletter
\providecommand \@ifxundefined [1]{%
 \@ifx{#1\undefined}
}%
\providecommand \@ifnum [1]{%
 \ifnum #1\expandafter \@firstoftwo
 \else \expandafter \@secondoftwo
 \fi
}%
\providecommand \@ifx [1]{%
 \ifx #1\expandafter \@firstoftwo
 \else \expandafter \@secondoftwo
 \fi
}%
\providecommand \natexlab [1]{#1}%
\providecommand \enquote  [1]{``#1''}%
\providecommand \bibnamefont  [1]{#1}%
\providecommand \bibfnamefont [1]{#1}%
\providecommand \citenamefont [1]{#1}%
\providecommand \href@noop [0]{\@secondoftwo}%
\providecommand \href [0]{\begingroup \@sanitize@url \@href}%
\providecommand \@href[1]{\@@startlink{#1}\@@href}%
\providecommand \@@href[1]{\endgroup#1\@@endlink}%
\providecommand \@sanitize@url [0]{\catcode `\\12\catcode `\$12\catcode
  `\&12\catcode `\#12\catcode `\^12\catcode `\_12\catcode `\%12\relax}%
\providecommand \@@startlink[1]{}%
\providecommand \@@endlink[0]{}%
\providecommand \url  [0]{\begingroup\@sanitize@url \@url }%
\providecommand \@url [1]{\endgroup\@href {#1}{\urlprefix }}%
\providecommand \urlprefix  [0]{URL }%
\providecommand \Eprint [0]{\href }%
\providecommand \doibase [0]{http://dx.doi.org/}%
\providecommand \selectlanguage [0]{\@gobble}%
\providecommand \bibinfo  [0]{\@secondoftwo}%
\providecommand \bibfield  [0]{\@secondoftwo}%
\providecommand \translation [1]{[#1]}%
\providecommand \BibitemOpen [0]{}%
\providecommand \bibitemStop [0]{}%
\providecommand \bibitemNoStop [0]{.\EOS\space}%
\providecommand \EOS [0]{\spacefactor3000\relax}%
\providecommand \BibitemShut  [1]{\csname bibitem#1\endcsname}%
\let\auto@bib@innerbib\@empty
\bibitem [{\citenamefont {Kohn}\ and\ \citenamefont
  {Sham}(1965)}]{Kohn65pA1133}%
  \BibitemOpen
  \bibfield  {author} {\bibinfo {author} {\bibfnamefont {W.}~\bibnamefont
  {Kohn}}\ and\ \bibinfo {author} {\bibfnamefont {L.~J.}\ \bibnamefont
  {Sham}},\ }\href@noop {} {\bibfield  {journal} {\bibinfo  {journal} {Phys.
  Rev.}\ }\textbf {\bibinfo {volume} {140}},\ \bibinfo {pages} {A1133}
  (\bibinfo {year} {1965})}\BibitemShut {NoStop}%
\bibitem [{\citenamefont {Perdew}\ \emph
  {et~al.}(1996{\natexlab{a}})\citenamefont {Perdew}, \citenamefont {Burke},\
  and\ \citenamefont {Ernzerhof}}]{Perdew96p3865}%
  \BibitemOpen
  \bibfield  {author} {\bibinfo {author} {\bibfnamefont {J.~P.}\ \bibnamefont
  {Perdew}}, \bibinfo {author} {\bibfnamefont {K.}~\bibnamefont {Burke}}, \
  and\ \bibinfo {author} {\bibfnamefont {M.}~\bibnamefont {Ernzerhof}},\
  }\href@noop {} {\bibfield  {journal} {\bibinfo  {journal} {Phys. Rev. Lett.}\
  }\textbf {\bibinfo {volume} {77}},\ \bibinfo {pages} {3865} (\bibinfo {year}
  {1996}{\natexlab{a}})}\BibitemShut {NoStop}%
\bibitem [{\citenamefont {Bühl}\ \emph {et~al.}(2008)\citenamefont {Bühl},
  \citenamefont {Reimann}, \citenamefont {Pantazis}, \citenamefont {Bredow},\
  and\ \citenamefont {Neese}}]{Buhl08p1449}%
  \BibitemOpen
  \bibfield  {author} {\bibinfo {author} {\bibfnamefont {M.}~\bibnamefont
  {Bühl}}, \bibinfo {author} {\bibfnamefont {C.}~\bibnamefont {Reimann}},
  \bibinfo {author} {\bibfnamefont {D.~A.}\ \bibnamefont {Pantazis}}, \bibinfo
  {author} {\bibfnamefont {T.}~\bibnamefont {Bredow}}, \ and\ \bibinfo {author}
  {\bibfnamefont {F.}~\bibnamefont {Neese}},\ }\href@noop {} {\bibfield
  {journal} {\bibinfo  {journal} {Journal of chemical theory and computation}\
  }\textbf {\bibinfo {volume} {4}},\ \bibinfo {pages} {1449} (\bibinfo {year}
  {2008})}\BibitemShut {NoStop}%
\bibitem [{\citenamefont {Perdew}(1985)}]{Perdew85p497}%
  \BibitemOpen
  \bibfield  {author} {\bibinfo {author} {\bibfnamefont {J.~P.}\ \bibnamefont
  {Perdew}},\ }\href@noop {} {\bibfield  {journal} {\bibinfo  {journal}
  {International Journal of Quantum Chemistry}\ }\textbf {\bibinfo {volume}
  {28}},\ \bibinfo {pages} {497} (\bibinfo {year} {1985})}\BibitemShut
  {NoStop}%
\bibitem [{\citenamefont {Adamo}\ and\ \citenamefont
  {Barone}(1998)}]{Adamo99p6158}%
  \BibitemOpen
  \bibfield  {author} {\bibinfo {author} {\bibfnamefont {C.}~\bibnamefont
  {Adamo}}\ and\ \bibinfo {author} {\bibfnamefont {V.}~\bibnamefont {Barone}},\
  }\href@noop {} {\bibfield  {journal} {\bibinfo  {journal} {J. Chem. Phys.}\
  }\textbf {\bibinfo {volume} {110}},\ \bibinfo {pages} {6158} (\bibinfo {year}
  {1998})}\BibitemShut {NoStop}%
\bibitem [{\citenamefont {Muscat}\ \emph {et~al.}(2001)\citenamefont {Muscat},
  \citenamefont {Wander},\ and\ \citenamefont {Harrison}}]{Muscat01p397}%
  \BibitemOpen
  \bibfield  {author} {\bibinfo {author} {\bibfnamefont {J.}~\bibnamefont
  {Muscat}}, \bibinfo {author} {\bibfnamefont {A.}~\bibnamefont {Wander}}, \
  and\ \bibinfo {author} {\bibfnamefont {N.~M.}\ \bibnamefont {Harrison}},\
  }\href@noop {} {\bibfield  {journal} {\bibinfo  {journal} {Chem. Phys.
  Lett.}\ }\textbf {\bibinfo {volume} {342}},\ \bibinfo {pages} {397} (\bibinfo
  {year} {2001})}\BibitemShut {NoStop}%
\bibitem [{\citenamefont {Heyd}\ \emph {et~al.}(2005)\citenamefont {Heyd},
  \citenamefont {Peralta}, \citenamefont {Scuseria},\ and\ \citenamefont
  {Martin}}]{Heyd05p174101}%
  \BibitemOpen
  \bibfield  {author} {\bibinfo {author} {\bibfnamefont {J.}~\bibnamefont
  {Heyd}}, \bibinfo {author} {\bibfnamefont {J.~E.}\ \bibnamefont {Peralta}},
  \bibinfo {author} {\bibfnamefont {G.~E.}\ \bibnamefont {Scuseria}}, \ and\
  \bibinfo {author} {\bibfnamefont {R.~L.}\ \bibnamefont {Martin}},\
  }\href@noop {} {\bibfield  {journal} {\bibinfo  {journal} {The Journal of
  chemical physics}\ }\textbf {\bibinfo {volume} {123}},\ \bibinfo {pages}
  {174101} (\bibinfo {year} {2005})}\BibitemShut {NoStop}%
\bibitem [{\citenamefont {Rappe}\ \emph {et~al.}(1990)\citenamefont {Rappe},
  \citenamefont {Rabe}, \citenamefont {Kaxiras},\ and\ \citenamefont
  {Joannopoulos}}]{Rappe90p1227}%
  \BibitemOpen
  \bibfield  {author} {\bibinfo {author} {\bibfnamefont {A.~M.}\ \bibnamefont
  {Rappe}}, \bibinfo {author} {\bibfnamefont {K.~M.}\ \bibnamefont {Rabe}},
  \bibinfo {author} {\bibfnamefont {E.}~\bibnamefont {Kaxiras}}, \ and\
  \bibinfo {author} {\bibfnamefont {J.~D.}\ \bibnamefont {Joannopoulos}},\
  }\href@noop {} {\bibfield  {journal} {\bibinfo  {journal} {Phys. Rev. B Rapid
  Comm.}\ }\textbf {\bibinfo {volume} {41}},\ \bibinfo {pages} {1227} (\bibinfo
  {year} {1990})}\BibitemShut {NoStop}%
\bibitem [{\citenamefont {Ramer}\ and\ \citenamefont
  {Rappe}(1999)}]{Ramer99p12471}%
  \BibitemOpen
  \bibfield  {author} {\bibinfo {author} {\bibfnamefont {N.~J.}\ \bibnamefont
  {Ramer}}\ and\ \bibinfo {author} {\bibfnamefont {A.~M.}\ \bibnamefont
  {Rappe}},\ }\href@noop {} {\bibfield  {journal} {\bibinfo  {journal} {Phys.
  Rev. B}\ }\textbf {\bibinfo {volume} {59}},\ \bibinfo {pages} {12471}
  (\bibinfo {year} {1999})}\BibitemShut {NoStop}%
\bibitem [{\citenamefont {Perdew}\ and\ \citenamefont
  {Schmidt}(2001)}]{Perdew01p1}%
  \BibitemOpen
  \bibfield  {author} {\bibinfo {author} {\bibfnamefont {J.~P.}\ \bibnamefont
  {Perdew}}\ and\ \bibinfo {author} {\bibfnamefont {K.}~\bibnamefont
  {Schmidt}},\ }in\ \href@noop {} {\emph {\bibinfo {booktitle} {AIP Conference
  Proceedings}}}\ (\bibinfo {organization} {IOP INSTITUTE OF PHYSICS PUBLISHING
  LTD},\ \bibinfo {year} {2001})\ pp.\ \bibinfo {pages} {1--20}\BibitemShut
  {NoStop}%
\bibitem [{\citenamefont {Fuchs}\ \emph {et~al.}(1998)\citenamefont {Fuchs},
  \citenamefont {Bockstedte}, \citenamefont {Pehlke},\ and\ \citenamefont
  {Scheffler}}]{Fuchs98p2134}%
  \BibitemOpen
  \bibfield  {author} {\bibinfo {author} {\bibfnamefont {M.}~\bibnamefont
  {Fuchs}}, \bibinfo {author} {\bibfnamefont {M.}~\bibnamefont {Bockstedte}},
  \bibinfo {author} {\bibfnamefont {E.}~\bibnamefont {Pehlke}}, \ and\ \bibinfo
  {author} {\bibfnamefont {M.}~\bibnamefont {Scheffler}},\ }\href@noop {}
  {\bibfield  {journal} {\bibinfo  {journal} {Physical Review B}\ }\textbf
  {\bibinfo {volume} {57}},\ \bibinfo {pages} {2134} (\bibinfo {year}
  {1998})}\BibitemShut {NoStop}%
\bibitem [{\citenamefont {Trail}\ and\ \citenamefont
  {Needs}(2005)}]{Trail05p014112}%
  \BibitemOpen
  \bibfield  {author} {\bibinfo {author} {\bibfnamefont {J.~R.}\ \bibnamefont
  {Trail}}\ and\ \bibinfo {author} {\bibfnamefont {R.~J.}\ \bibnamefont
  {Needs}},\ }\href@noop {} {\bibfield  {journal} {\bibinfo  {journal} {J.
  Chem. Phys.}\ }\textbf {\bibinfo {volume} {122}},\ \bibinfo {pages} {014112}
  (\bibinfo {year} {2005})}\BibitemShut {NoStop}%
\bibitem [{\citenamefont {Al-Saidi}\ \emph {et~al.}(2008)\citenamefont
  {Al-Saidi}, \citenamefont {Walter},\ and\ \citenamefont
  {Rappe}}]{Al-Saidi08p075112}%
  \BibitemOpen
  \bibfield  {author} {\bibinfo {author} {\bibfnamefont {W.}~\bibnamefont
  {Al-Saidi}}, \bibinfo {author} {\bibfnamefont {E.~J.}\ \bibnamefont
  {Walter}}, \ and\ \bibinfo {author} {\bibfnamefont {A.~M.}\ \bibnamefont
  {Rappe}},\ }\href@noop {} {\bibfield  {journal} {\bibinfo  {journal} {Phys.
  Rev. B}\ }\textbf {\bibinfo {volume} {77}},\ \bibinfo {pages} {075112}
  (\bibinfo {year} {2008})}\BibitemShut {NoStop}%
\bibitem [{\citenamefont {Greeff}\ \emph {et~al.}(1998)\citenamefont {Greeff},
  \citenamefont {Lester} \emph {et~al.}}]{Greeff98p1607}%
  \BibitemOpen
  \bibfield  {author} {\bibinfo {author} {\bibfnamefont {C.}~\bibnamefont
  {Greeff}}, \bibinfo {author} {\bibfnamefont {W.}~\bibnamefont {Lester}},
  \emph {et~al.},\ }\href@noop {} {\bibfield  {journal} {\bibinfo  {journal}
  {The Journal of chemical physics}\ }\textbf {\bibinfo {volume} {109}},\
  \bibinfo {pages} {1607} (\bibinfo {year} {1998})}\BibitemShut {NoStop}%
\bibitem [{\citenamefont {Ovcharenko}\ \emph {et~al.}(2001)\citenamefont
  {Ovcharenko}, \citenamefont {Aspuru-Guzik},\ and\ \citenamefont
  {Lester~Jr}}]{Ovcharenko01p7790}%
  \BibitemOpen
  \bibfield  {author} {\bibinfo {author} {\bibfnamefont {I.}~\bibnamefont
  {Ovcharenko}}, \bibinfo {author} {\bibfnamefont {A.}~\bibnamefont
  {Aspuru-Guzik}}, \ and\ \bibinfo {author} {\bibfnamefont {W.~A.}\
  \bibnamefont {Lester~Jr}},\ }\href@noop {} {\bibfield  {journal} {\bibinfo
  {journal} {The Journal of Chemical Physics}\ }\textbf {\bibinfo {volume}
  {114}},\ \bibinfo {pages} {7790} (\bibinfo {year} {2001})}\BibitemShut
  {NoStop}%
\bibitem [{\citenamefont {Wu}\ \emph {et~al.}(2009)\citenamefont {Wu},
  \citenamefont {Walter}, \citenamefont {Rappe}, \citenamefont {Car},\ and\
  \citenamefont {Selloni}}]{Wu09p115201}%
  \BibitemOpen
  \bibfield  {author} {\bibinfo {author} {\bibfnamefont {X.~F.}\ \bibnamefont
  {Wu}}, \bibinfo {author} {\bibfnamefont {E.~J.}\ \bibnamefont {Walter}},
  \bibinfo {author} {\bibfnamefont {A.~M.}\ \bibnamefont {Rappe}}, \bibinfo
  {author} {\bibfnamefont {R.}~\bibnamefont {Car}}, \ and\ \bibinfo {author}
  {\bibfnamefont {A.}~\bibnamefont {Selloni}},\ }\href@noop {} {\bibfield
  {journal} {\bibinfo  {journal} {Phys. Rev. B}\ }\textbf {\bibinfo {volume}
  {80}},\ \bibinfo {pages} {115201} (\bibinfo {year} {2009})}\BibitemShut
  {NoStop}%
\bibitem [{Opi()}]{Opium}%
  \BibitemOpen
  \href@noop {} {}\bibinfo {howpublished}
  {http://opium.sourceforge.net}\BibitemShut {NoStop}%
\bibitem [{\citenamefont {Hamann}\ \emph {et~al.}(1979)\citenamefont {Hamann},
  \citenamefont {Schl\"uter},\ and\ \citenamefont {Chiang}}]{Hamann79p1494}%
  \BibitemOpen
  \bibfield  {author} {\bibinfo {author} {\bibfnamefont {D.~R.}\ \bibnamefont
  {Hamann}}, \bibinfo {author} {\bibfnamefont {M.}~\bibnamefont {Schl\"uter}},
  \ and\ \bibinfo {author} {\bibfnamefont {C.}~\bibnamefont {Chiang}},\
  }\href@noop {} {\bibfield  {journal} {\bibinfo  {journal} {Phys. Rev. Lett.}\
  }\textbf {\bibinfo {volume} {43}},\ \bibinfo {pages} {1494} (\bibinfo {year}
  {1979})}\BibitemShut {NoStop}%
\bibitem [{\citenamefont {Kleinman}\ and\ \citenamefont
  {Bylander}(1982)}]{Kleinman82p1425}%
  \BibitemOpen
  \bibfield  {author} {\bibinfo {author} {\bibfnamefont {L.}~\bibnamefont
  {Kleinman}}\ and\ \bibinfo {author} {\bibfnamefont {D.~M.}\ \bibnamefont
  {Bylander}},\ }\href@noop {} {\bibfield  {journal} {\bibinfo  {journal}
  {Phys. Rev. Lett.}\ }\textbf {\bibinfo {volume} {48}},\ \bibinfo {pages}
  {1425} (\bibinfo {year} {1982})}\BibitemShut {NoStop}%
\bibitem [{\citenamefont {Fischer}(1987)}]{Fischer87p355}%
  \BibitemOpen
  \bibfield  {author} {\bibinfo {author} {\bibfnamefont {C.~F.}\ \bibnamefont
  {Fischer}},\ }\href@noop {} {\bibfield  {journal} {\bibinfo  {journal}
  {Computer physics communications}\ }\textbf {\bibinfo {volume} {43}},\
  \bibinfo {pages} {355} (\bibinfo {year} {1987})}\BibitemShut {NoStop}%
\bibitem [{\citenamefont {Slater}(1960)}]{Slater}%
  \BibitemOpen
  \bibfield  {author} {\bibinfo {author} {\bibfnamefont {J.~C.}\ \bibnamefont
  {Slater}},\ }\href@noop {} {\emph {\bibinfo {title} {Quantum Theory of Atomic
  Structure}}}\ (\bibinfo  {publisher} {McGraw-Hill Book Company, Inc.},\
  \bibinfo {year} {1960})\BibitemShut {NoStop}%
\bibitem [{\citenamefont {Fischer}(1977)}]{Fischer}%
  \BibitemOpen
  \bibfield  {author} {\bibinfo {author} {\bibfnamefont {C.~F.}\ \bibnamefont
  {Fischer}},\ }\href@noop {} {\emph {\bibinfo {title} {The Hartree--Fock
  method for Atoms: a numerical approach}}}\ (\bibinfo  {publisher} {A
  Wiley-Interscience publication. New York},\ \bibinfo {year}
  {1977})\BibitemShut {NoStop}%
\bibitem [{\citenamefont {Perdew}\ \emph
  {et~al.}(1996{\natexlab{b}})\citenamefont {Perdew}, \citenamefont
  {Ernzerhof},\ and\ \citenamefont {Burke}}]{Perdew96p9982}%
  \BibitemOpen
  \bibfield  {author} {\bibinfo {author} {\bibfnamefont {J.~P.}\ \bibnamefont
  {Perdew}}, \bibinfo {author} {\bibfnamefont {M.}~\bibnamefont {Ernzerhof}}, \
  and\ \bibinfo {author} {\bibfnamefont {K.}~\bibnamefont {Burke}},\
  }\href@noop {} {\bibfield  {journal} {\bibinfo  {journal} {J. Chem. Phys.}\
  }\textbf {\bibinfo {volume} {105}},\ \bibinfo {pages} {9982} (\bibinfo {year}
  {1996}{\natexlab{b}})}\BibitemShut {NoStop}%
\bibitem [{\citenamefont {Giannozzi}\ \emph {et~al.}(2009)\citenamefont
  {Giannozzi}, \citenamefont {Baroni}, \citenamefont {Bonini}, \citenamefont
  {Calandra}, \citenamefont {Car}, \citenamefont {Cavazzoni}, \citenamefont
  {Ceresoli}, \citenamefont {Chiarotti}, \citenamefont {Cococcioni},
  \citenamefont {Dabo}, \citenamefont {Corso}, \citenamefont {de~Gironcoli},
  \citenamefont {Fabris}, \citenamefont {Fratesi}, \citenamefont {Gebauer},
  \citenamefont {Gerstmann}, \citenamefont {Gougoussis}, \citenamefont
  {Kokalj}, \citenamefont {Lazzeri}, \citenamefont {Martin-Samos},
  \citenamefont {Marzari}, \citenamefont {Mauri}, \citenamefont {Mazzarello},
  \citenamefont {Paolini}, \citenamefont {Pasquarello}, \citenamefont
  {Paulatto}, \citenamefont {Sbraccia}, \citenamefont {Scandolo}, \citenamefont
  {Sclauzero}, \citenamefont {Seitsonen}, \citenamefont {Smogunov},
  \citenamefont {Umari},\ and\ \citenamefont
  {Wentzcovitch}}]{Giannozzi09p395502}%
  \BibitemOpen
  \bibfield  {author} {\bibinfo {author} {\bibfnamefont {P.}~\bibnamefont
  {Giannozzi}}, \bibinfo {author} {\bibfnamefont {S.}~\bibnamefont {Baroni}},
  \bibinfo {author} {\bibfnamefont {N.}~\bibnamefont {Bonini}}, \bibinfo
  {author} {\bibfnamefont {M.}~\bibnamefont {Calandra}}, \bibinfo {author}
  {\bibfnamefont {R.}~\bibnamefont {Car}}, \bibinfo {author} {\bibfnamefont
  {C.}~\bibnamefont {Cavazzoni}}, \bibinfo {author} {\bibfnamefont
  {D.}~\bibnamefont {Ceresoli}}, \bibinfo {author} {\bibfnamefont {G.~L.}\
  \bibnamefont {Chiarotti}}, \bibinfo {author} {\bibfnamefont {M.}~\bibnamefont
  {Cococcioni}}, \bibinfo {author} {\bibfnamefont {I.}~\bibnamefont {Dabo}},
  \bibinfo {author} {\bibfnamefont {A.~D.}\ \bibnamefont {Corso}}, \bibinfo
  {author} {\bibfnamefont {S.}~\bibnamefont {de~Gironcoli}}, \bibinfo {author}
  {\bibfnamefont {S.}~\bibnamefont {Fabris}}, \bibinfo {author} {\bibfnamefont
  {G.}~\bibnamefont {Fratesi}}, \bibinfo {author} {\bibfnamefont
  {R.}~\bibnamefont {Gebauer}}, \bibinfo {author} {\bibfnamefont
  {U.}~\bibnamefont {Gerstmann}}, \bibinfo {author} {\bibfnamefont
  {C.}~\bibnamefont {Gougoussis}}, \bibinfo {author} {\bibfnamefont
  {A.}~\bibnamefont {Kokalj}}, \bibinfo {author} {\bibfnamefont
  {M.}~\bibnamefont {Lazzeri}}, \bibinfo {author} {\bibfnamefont
  {L.}~\bibnamefont {Martin-Samos}}, \bibinfo {author} {\bibfnamefont
  {N.}~\bibnamefont {Marzari}}, \bibinfo {author} {\bibfnamefont
  {F.}~\bibnamefont {Mauri}}, \bibinfo {author} {\bibfnamefont
  {R.}~\bibnamefont {Mazzarello}}, \bibinfo {author} {\bibfnamefont
  {S.}~\bibnamefont {Paolini}}, \bibinfo {author} {\bibfnamefont
  {A.}~\bibnamefont {Pasquarello}}, \bibinfo {author} {\bibfnamefont
  {L.}~\bibnamefont {Paulatto}}, \bibinfo {author} {\bibfnamefont
  {C.}~\bibnamefont {Sbraccia}}, \bibinfo {author} {\bibfnamefont
  {S.}~\bibnamefont {Scandolo}}, \bibinfo {author} {\bibfnamefont
  {G.}~\bibnamefont {Sclauzero}}, \bibinfo {author} {\bibfnamefont {A.~P.}\
  \bibnamefont {Seitsonen}}, \bibinfo {author} {\bibfnamefont {A.}~\bibnamefont
  {Smogunov}}, \bibinfo {author} {\bibfnamefont {P.}~\bibnamefont {Umari}}, \
  and\ \bibinfo {author} {\bibfnamefont {R.~M.}\ \bibnamefont {Wentzcovitch}},\
  }\href@noop {} {\bibfield  {journal} {\bibinfo  {journal} {J. Phys.: Condens.
  Matter}\ }\textbf {\bibinfo {volume} {21}},\ \bibinfo {pages} {395502}
  (\bibinfo {year} {2009})}\BibitemShut {NoStop}%
\bibitem [{\citenamefont {Blum}\ \emph {et~al.}(2009)\citenamefont {Blum},
  \citenamefont {Gehrke}, \citenamefont {Hanke}, \citenamefont {Havu},
  \citenamefont {Havu}, \citenamefont {Ren}, \citenamefont {Reuter},\ and\
  \citenamefont {Scheffler}}]{Blum09p2175}%
  \BibitemOpen
  \bibfield  {author} {\bibinfo {author} {\bibfnamefont {V.}~\bibnamefont
  {Blum}}, \bibinfo {author} {\bibfnamefont {R.}~\bibnamefont {Gehrke}},
  \bibinfo {author} {\bibfnamefont {F.}~\bibnamefont {Hanke}}, \bibinfo
  {author} {\bibfnamefont {P.}~\bibnamefont {Havu}}, \bibinfo {author}
  {\bibfnamefont {V.}~\bibnamefont {Havu}}, \bibinfo {author} {\bibfnamefont
  {X.}~\bibnamefont {Ren}}, \bibinfo {author} {\bibfnamefont {K.}~\bibnamefont
  {Reuter}}, \ and\ \bibinfo {author} {\bibfnamefont {M.}~\bibnamefont
  {Scheffler}},\ }\href {http://www.fhi-berlin.mpg.de/aims/} {\bibfield
  {journal} {\bibinfo  {journal} {Computer Physics Communications}\ }\textbf
  {\bibinfo {volume} {180}},\ \bibinfo {pages} {2175} (\bibinfo {year}
  {2009})}\BibitemShut {NoStop}%
\bibitem [{\citenamefont {Becke}(1998)}]{Becke98p2092}%
  \BibitemOpen
  \bibfield  {author} {\bibinfo {author} {\bibfnamefont {A.~D.}\ \bibnamefont
  {Becke}},\ }\href@noop {} {\bibfield  {journal} {\bibinfo  {journal} {The
  Journal of chemical physics}\ }\textbf {\bibinfo {volume} {109}},\ \bibinfo
  {pages} {2092} (\bibinfo {year} {1998})}\BibitemShut {NoStop}%
\bibitem [{\citenamefont {Ernzerhof}\ and\ \citenamefont
  {Scuseria}(1999)}]{Ernzerhof99p5029}%
  \BibitemOpen
  \bibfield  {author} {\bibinfo {author} {\bibfnamefont {M.}~\bibnamefont
  {Ernzerhof}}\ and\ \bibinfo {author} {\bibfnamefont {E.}~\bibnamefont
  {Scuseria}},\ }\href@noop {} {\bibfield  {journal} {\bibinfo  {journal} {J.
  Chem. Phys.}\ }\textbf {\bibinfo {volume} {110}},\ \bibinfo {pages} {5029}
  (\bibinfo {year} {1999})}\BibitemShut {NoStop}%
\bibitem [{\citenamefont {Matsushita}\ \emph {et~al.}(2011)\citenamefont
  {Matsushita}, \citenamefont {Nakamura},\ and\ \citenamefont
  {Oshiyama}}]{Matsushita11p075205}%
  \BibitemOpen
  \bibfield  {author} {\bibinfo {author} {\bibfnamefont {Y.-i.}\ \bibnamefont
  {Matsushita}}, \bibinfo {author} {\bibfnamefont {K.}~\bibnamefont
  {Nakamura}}, \ and\ \bibinfo {author} {\bibfnamefont {A.}~\bibnamefont
  {Oshiyama}},\ }\href@noop {} {\bibfield  {journal} {\bibinfo  {journal}
  {Physical Review B}\ }\textbf {\bibinfo {volume} {84}},\ \bibinfo {pages}
  {075205} (\bibinfo {year} {2011})}\BibitemShut {NoStop}%
\bibitem [{\citenamefont {Toulouse}\ \emph {et~al.}(2010)\citenamefont
  {Toulouse}, \citenamefont {Zhu}, \citenamefont {Angy{\'a}n},\ and\
  \citenamefont {Savin}}]{Toulouse10p032502}%
  \BibitemOpen
  \bibfield  {author} {\bibinfo {author} {\bibfnamefont {J.}~\bibnamefont
  {Toulouse}}, \bibinfo {author} {\bibfnamefont {W.}~\bibnamefont {Zhu}},
  \bibinfo {author} {\bibfnamefont {J.~G.}\ \bibnamefont {Angy{\'a}n}}, \ and\
  \bibinfo {author} {\bibfnamefont {A.}~\bibnamefont {Savin}},\ }\href@noop {}
  {\bibfield  {journal} {\bibinfo  {journal} {Physical Review A}\ }\textbf
  {\bibinfo {volume} {82}},\ \bibinfo {pages} {032502} (\bibinfo {year}
  {2010})}\BibitemShut {NoStop}%
\end{thebibliography}%
\end{document}